\documentclass[ aps,%
floatfix,%
oneside,%
onecolumn,%
nobibnotes,%
nofootinbib,%
superscriptaddress,%
showpacs,%
]%
{revtex4}
\usepackage{hyperref}

\newcommand{\A}{\mathcal{A}}

\newcommand{\phiL}{\phi_\Vert}\newcommand{\phiT}{\phi_\perp}
\newcommand{\fL}{f} \newcommand{\fT}{f_\perp}

\newcommand{\JP}{\psi} \newcommand{\ec}{\eta_c}
\newcommand{\epem}{e^+e^-} 
\newcommand{\om}{\omega}

\newcommand{\MeV}{\,\mathrm{MeV}}
\newcommand{\GeV}{\,\mathrm{GeV}}

\newcommand{\Br}{\mathrm{Br}}

\newcommand{\beq}{\begin{eqnarray}}\newcommand{\eeq}{\end{eqnarray}}
\newcommand{\beqa}{\begin{eqnarray*}}\newcommand{\eeqa}{\end{eqnarray*}}

\begin{document}

\title{Leading twist contribution to color singlet $\chi_{c0,2}\to\omega\omega$ decays}
\author{A.V. Luchinsky}
\email{Alexey.Luchinsky@ihep.ru}
\affiliation{Institute for High Energy Physics, Protvino, Russia}
\begin{abstract}
In this paper the leading twist contribution to $\chi_{c0,2}\to\omega\omega$ decays in the color singlet approximation is considered. It is shown, that the predictions for $\Br(\chi_0\to\om\om)$ is in a good agreement with the experimental data, while $\Br(\chi_{c2}\to\om\om)$ differs from the experiment significantly.
\end{abstract}

\maketitle

\section{Introduction}

Recently the discrepancy between theoretical prediction for $\JP=J/\psi$ and $\ec$ production in $\epem$ annihilation at $\sqrt{s}=10.6\GeV$ \cite{Braaten:2002fi} and experimental result \cite{Abe:2002rb} has found a surprisingly simple explanation. In the works \cite{Ma:2004qf,Bondar:2004sv} it was shown that taking into account the intrinsic motion of quarks inside the charmonium mesons in the hard part of the amplitude one can significantly increase the theoretical predictions for the cross section of this reaction and reach the agreement with the experiment.

In the recent work \cite{Braguta:2005gw} we have confirmed this result using a slightly different model. In that paper we have also studied the influence of internal quark motion on scalar and tensor mesons decays into two vector ones (that is $\chi_{0,2}\to VV$. Specifically the decays $\chi_{c0,2}\to\rho\rho,\phi\phi$ and $\chi_{b0,2}\to\JP\JP$ were considered) and shown that the branching fractions of these decays also increase when one takes into consideration in intrinsic quark motion. For example, the agreement between theoretical predictions and experimental result for the $\chi_{c0}\to\phi\phi$ branching fraction can be reached. We have also studied the possibility of using $\chi_{b0,2}\to\JP\JP$ decay for $\chi_{b0,2}$ mesons observation at Tevatron and LHC colliders.

In the recent paper the \cite{BES:2005ds} $\chi_{c0,2}$ mesons were observed in the $\omega\omega$ mode and the results for the branching fractions $\Br(\chi_{c0,2}\to\om\om)$ were presented. The aim of this short note is to use the formulae presented  in \cite{Braguta:2005gw} for these decays and to compare our results with the experimental ones.

\section{Helicity matrix elements and distribution functions}

Analytical formulae for the width of the decay $\chi_0\to VV$ were presented in the work \cite{
Anselmino:1992rw} and we
will use this results in our paper. The nonzero helicity amplitudes
$\A^{(0)}_{\lambda_1,\lambda_2}$ of the decay of
scalar meson $\chi_0$ into vector mesons $V_1$ and $V_2$ with the helicities $\lambda_1$ and
$\lambda_2$ are given by the
expressions
\beqa
\A_{1,1}^{(0)} & = &  \A_{-1,-1}^{(0)}=
  i\frac{2^{13}}{9\sqrt{3}}\pi^3\alpha_s^2\epsilon^2
  \frac{|R'(0)|}{M^{4}}\fT^2I_{1,1}^{(0)}(\epsilon), \label{eq:A11}\\ %
\A_{0,0}^{(0)} & = &
  -i\frac{2^{12}}{9\sqrt{3}}\pi^3\alpha_s^2
  \frac{|R'(0)|}{M}\fL^2I_{0,0}^{(0)}(\epsilon), \label{eq:A00}
\eeqa
where
\beqa
  \epsilon & = & m/M,
\eeqa
$m$ and $M$ are masses of vector and scalar mesons respectively, $R(r)$ is the radial part of the
scalar meson's wave
function, $\fL$ and $\fT$ are longitudinal and transverse leptonic constants of the vector meson and
coefficients
$I^{(0)}_{\lambda_1,\lambda_2}$ are equal to
\beqa
  I_{1,1}^{(0)} & = &
    -\frac{1}{32}\int\limits _{0}^{1}dx\, dy \phiT(x)\phiT(y)
    \frac{1}{xy+(x-y)^2\epsilon^2}
    \frac{1}{(1-x)(1-y)+(x-y)^2\epsilon^2} \times
\nonumber \\ & \times &
   \frac{1}{2xy-x-y+2(x-y)^2\epsilon^2}
   \left[
     1+\frac{1}{2}\frac{(x-y)^2(1-4\epsilon^2)}
     {2xy-x-y+2(x-y)^2\epsilon^2}
   \right],\label{eq:I11}
\\
  I_{0,0}^{(0)} & = &
    -\frac{1}{32}\int\limits _0^1 dx\, dy \phiL(x)\phiL(y)
    \frac{1}{xy+(x-y)^2\epsilon^2}
    \frac{1}{1-x)(1-y)+(x-y)^2\epsilon^2} \times
\\  & \times &
    \frac{1}{2xy-x-y+2(x-y)^2\epsilon^2}
    \left\{
      1-\frac{1}{2}\frac{(x-y)^2(1-4\epsilon^2)}{2xy-x-y+2(x-y)^2\epsilon^2}-
\right.\nonumber \\  & - & \left.
    -2\epsilon^{2}\left[
       1+\frac{1}{2}\frac{(x-y)^2(1-4\epsilon^2)}{2xy-x-y+2(x-y)^2\epsilon^2}
     \right]\right\} .\label{eq:I00}
\eeqa
In the above equations $x$ and $y$ are the momentum fractions of the final mesons, carried by
quarks and
$\phi_{\Vert,\perp}(x)$ are longitudinal and transverse distribution functions of these quarks in mesons.

In \cite{Anselmino:1992rw} the similar formulae for nonzero helicity amplitudes of tensor meson
decay are also presented:
\beqa
  \A^{(2)}_{\lambda_1\lambda_2;\mu} & = &
    \tilde\A_{\lambda_1\lambda_2}e^{i\mu\varphi}d^{(2)}_{\mu,\lambda_1-\lambda_1}(\theta),
\eeqa
where $\mu$ is the meson spin projection on fixed axe, $\theta$ and $\varphi$ are polar and
azimuthal angles of one
of the final mesons in $\chi_2$ rest frame and reduced amplitudes $\tilde\A_{\lambda_1\lambda_2}$
are given by the expressions
\beqa
  \tilde\A_{1,1} & = & \tilde\A_{-1,-1} =
    -i \frac{2^{13}\sqrt{2}}{9\sqrt{3}}\pi^3\alpha_s^2\epsilon^2\frac{|R'(0)|}{M^4}\fT^2 I^{(2)}_
    {1,1} ,
  \\
  \tilde\A_{1,0} & = & \tilde\A_{0,1}=\tilde\A_{-1,0}=\tilde\A_{0,-1} =
    -i \frac{2^{12}\sqrt{2}}{9}\pi^3\alpha_s^2\epsilon\frac{|R'(0)|}{M^4}\fT \fL I^{(2)}_{1,0} ,
  \\
  \tilde\A_{1,-1} & = & \tilde\A_{-1,1} =
    -i \frac{2^{12}}{9}\pi^3\alpha_s^2\frac{|R'(0)|}{M^4}\fT^2 I^{(2)}_{1,-1} ,
  \\
  \tilde\A_{0,0} & = &
    -i \frac{2^{11}\sqrt{2}}{9\sqrt{3}}\pi^3\alpha_s^2\frac{|R'(0)|}{M^4}\fL^2 I^{(2)}_{0,0} ,
\\
  I^{(2)}_{1,1} & = & I^{(0)}_{1,1},
\\
  I^{(2)}_{1,0} & = &
    -\frac{1}{32}\int\limits_0^1\int\limits_0^1 dx dy \phiT(x)\phiL(y)
    \frac{1}{xy+(x-y)^2\epsilon^2} \frac{1}{(1-x)(1-y)+(x-y)^2\epsilon^2}
  \times \\ & \times &
    \frac{1}{2xy-x-y+2(x-y)^2\epsilon^2} \left[
      1+\frac{1}{2}\frac{(x-y)^2(1-4\epsilon^2)}{2xy-x-y+2(x-y)^2\epsilon^2}
    \right] ,
\\
  I^{(2)}_{1,-1} & = &
    -\frac{1}{32}\int\limits_0^1\int\limits_0^1 dx dy \phiT(x)\phiL(y)
    \frac{1}{xy+(x-y)^2\epsilon^2} \frac{1}{(1-x)(1-y)+(x-y)^2\epsilon^2}
  \times \\ & \times &
    \frac{1}{2xy-x-y+2(x-y)^2\epsilon^2},
\\
  I^{(2)}_{0,0} & = &
    -\frac{1}{32}\int\limits_0^1\int\limits_0^1 dx dy \phiT(x)\phiL(y)
    \frac{1}{xy+(x-y)^2\epsilon^2} \frac{1}{(1-x)(1-y)+(x-y)^2\epsilon^2}
  \times \\ & \times &
  \frac{1}{2xy-x-y+2(x-y)^2\epsilon^2} \left\{
    1+\frac{(x-y)^2(1-4\epsilon^2)}{2xy-x-y+2(x-y)^2\epsilon^2}+ \right.
  \\ &+& \left.
    4\epsilon^2\left[
      1+\frac{1}{2}\frac{(x-y)^2(1-4\epsilon^2)}{2xy-x-y+2(x-y)^2\epsilon^2}
    \right] \right\}.
\eeqa

The leading twist structure functions $\phiL(x)$ and $\phiT(x)$ can be expressed through the Gegenbauer polynomials (see \cite{Lepage:1980fj} and refrences therein)
\beq
\phi_{\Vert,\perp}(x) & = & 6x(1-x)\left[ 
  1+\sum\limits_{n=2,4,\dots} a_n^{\Vert,\perp} C^{3/2}_n (2x-1)
\right].
\label{eq:phi}
\eeq
In what follows we will restrict ourself to first two terms of this expansion. The derivative of the $\chi_{c}$-meson wave function in the origin can be expressed through the decay widths of these mesons:
\beq
  \Gamma_{\chi_{c0}} & \approx &
    \Gamma(\chi_{c0}\to gg) = 96\frac{\alpha_s^2}{M_{\chi_{b0}}^4} |R'(0)|^2, 
    \label{eq:Rp0}\\
  \Gamma_{\chi_{c2}} & \approx &
    \Gamma(\chi_{c2}\to gg) = \frac{128}{5}\frac{\alpha_s^2}{M_{\chi_{b2}}^4} |R'(0)|^2 
    \label{eq:Rp2}
\eeq
and the longitudinal $\om$-meson leptonic constant can be expressed through the $\om\to\epem$ decay width using the relation
\beq
\Gamma(\om\to\epem) & = &
  \frac{4\pi}{3}\left(\frac{e_u+e_d}{\sqrt{2}}\right)^2 \alpha^2 \frac{\fL^2}{M_\om}.
  \label{eq:f}
\eeq
On the other hand, the derivation of the transverse leptonic constant $\fT$ and the structure function momenta $a_2^{\Vert,\perp}$ is not so simple. The values of these parameters can be obtained using QCD sum rules 
\cite{Chernyak:1983ej,Ball:1996tb} and will be discussed in the next section.

\section{Numerical results and conclusion}
With the help of equations (\ref{eq:Rp0}),(\ref{eq:Rp2}),(\ref{eq:f}) we get the following values:
 \beqa
   |R'(0)|^2 & = & 0.16\GeV^5, \qquad f=196\MeV.
 \eeqa
 The values of transverse leptonic constant $f$ and structure function momenta $a_2^{\Vert,\perp}$ were obtained in 
 \cite{Chernyak:1983ej} and reanalyzed in \cite{Ball:1996tb}. According to the last paper, the values of these constants at the renormalization scale $\mu_0=1\GeV$ are equal to
 \beqa
   \fT(\mu_0)=(160\pm 10)\MeV,\quad 
   a_2^\perp(\mu_0)&=&0.2\pm 0.1,\quad
   a_2^\Vert(\mu_0)=0.18\pm 0.10
 \eeqa
 The evaluation to other renormalization scale can be done using the equations
 \beqa
 a_2^\Vert(\mu)=a_2^\Vert(\mu_0)\left(
   \frac{\alpha_s(\mu)}{\alpha_s(\mu_0)}
 \right)^{2/3},\quad
 a_2^\perp(\mu)=a_2^\perp(\mu_0)\left(
   \frac{\alpha_s(\mu)}{\alpha_s(\mu_0)}
 \right)^{8/15},\quad
 \fT(\mu)=\fT(\mu_0)\left(
   \frac{\alpha_s(\mu)}{\alpha_s(\mu_0)}
 \right)^{4/27}.
 \eeqa
 Using these values of the parameters we obtain the following branching fractions:
 \beqa
 \Br(\chi_{c0}\to\om\om) & = & (2.3\pm 1.1)\cdot 10^{-3},\qquad 
 \Br(\chi_{c2}\to\om\om)   =   ( 6 \pm 3  )\cdot 10^{-3},
 \eeqa
 where the errors are caused by the errors in distribution functions momenta. In the massless quark approximation (that is using $\fT=\fL$, $a_2^\Vert=a_2^\perp=0$) we get
 \beqa
 \Br(\chi_{c0}\to\om\om) & = & 1.1\cdot 10^{-3},\qquad 
 \Br(\chi_{c2}\to\om\om)   =   5.7\cdot 10^{-3}.
 \eeqa
 These results should be compared with the experimental values
 \beqa
 \Br(\chi_{c0}\to\om\om) & = & (2.29\pm 0.58 \pm 0.41)\cdot 10^{-3},\qquad 
 \Br(\chi_{c2}\to\om\om)   =   (1.77\pm 0.47 \pm 0.36  )\cdot 10^{-3}.
 \eeqa

As it can be easily seen, the branching fractions of the $\chi_{c0,2}\to\om\om$ decays strongly depend on the choice of the meson structure functions and more precise values of their parameters is important. For example, the 50\% error in $\chi_{c2}\to\om\om$ branching fraction is mainly caused by the same errors in $a_2^{\Vert,\perp}$.

Our prediction for the $\chi_{c0}\to\om\om$ branching fraction is in excellent agreement with the experimental value. The $\Br(\chi_{c2}\to\om\om)$ result, on the contrary, differs significantly from the experiment. The reasons for such discrepantly could be poor knowledge of meson distribution functions or the contribution of the color-octet states that were neglected in this note.

\begin{acknowledgments}
The author thanks A.K. Likhoded and V.V. Braguta for useful discussions. This work was partially
supported by Russian Foundation of Basic Research under grant 04-02-17530, Russian Education
Ministry grant E02-31-96, CRDF grant MO-011-0, Scientific School grant SS-1303.2003.2.
\end{acknowledgments}

\bibliographystyle{apsrev}

\begin{thebibliography}{10}
\expandafter\ifx\csname natexlab\endcsname\relax\def\natexlab#1{#1}\fi
\expandafter\ifx\csname bibnamefont\endcsname\relax
  \def\bibnamefont#1{#1}\fi
\expandafter\ifx\csname bibfnamefont\endcsname\relax
  \def\bibfnamefont#1{#1}\fi
\expandafter\ifx\csname citenamefont\endcsname\relax
  \def\citenamefont#1{#1}\fi
\expandafter\ifx\csname url\endcsname\relax
  \def\url#1{\texttt{#1}}\fi
\expandafter\ifx\csname urlprefix\endcsname\relax\def\urlprefix{URL }\fi
\providecommand{\bibinfo}[2]{#2}
\providecommand{\eprint}[2][]{\url{#2}}

\bibitem[{\citenamefont{Braaten and Lee}(2003)}]{Braaten:2002fi}
\bibinfo{author}{\bibfnamefont{E.}~\bibnamefont{Braaten}} \bibnamefont{and}
  \bibinfo{author}{\bibfnamefont{J.}~\bibnamefont{Lee}},
  \bibinfo{journal}{Phys. Rev.} \textbf{\bibinfo{volume}{D67}},
  \bibinfo{pages}{054007} (\bibinfo{year}{2003}), \eprint{hep-ph/0211085}.

\bibitem[{\citenamefont{Abe et~al.}(2002)}]{Abe:2002rb}
\bibinfo{author}{\bibfnamefont{K.}~\bibnamefont{Abe}} \bibnamefont{et~al.}
  (\bibinfo{collaboration}{Belle}), \bibinfo{journal}{Phys. Rev. Lett.}
  \textbf{\bibinfo{volume}{89}}, \bibinfo{pages}{142001}
  (\bibinfo{year}{2002}), \eprint{hep-ex/0205104}.

\bibitem[{\citenamefont{Ma and Si}(2004)}]{Ma:2004qf}
\bibinfo{author}{\bibfnamefont{J.~P.} \bibnamefont{Ma}} \bibnamefont{and}
  \bibinfo{author}{\bibfnamefont{Z.~G.} \bibnamefont{Si}},
  \bibinfo{journal}{Phys. Rev.} \textbf{\bibinfo{volume}{D70}},
  \bibinfo{pages}{074007} (\bibinfo{year}{2004}), \eprint{hep-ph/0405111}.

\bibitem[{\citenamefont{Bondar and Chernyak}(2004)}]{Bondar:2004sv}
\bibinfo{author}{\bibfnamefont{A.~E.} \bibnamefont{Bondar}} \bibnamefont{and}
  \bibinfo{author}{\bibfnamefont{V.~L.} \bibnamefont{Chernyak}}
  (\bibinfo{year}{2004}), \eprint{hep-ph/0412335}.

\bibitem[{\citenamefont{Braguta et~al.}(2005)\citenamefont{Braguta, Likhoded,
  and Luchinsky}}]{Braguta:2005gw}
\bibinfo{author}{\bibfnamefont{V.~V.} \bibnamefont{Braguta}},
  \bibinfo{author}{\bibfnamefont{A.~K.} \bibnamefont{Likhoded}},
  \bibnamefont{and} \bibinfo{author}{\bibfnamefont{A.~V.}
  \bibnamefont{Luchinsky}} (\bibinfo{year}{2005}), \eprint{hep-ph/0506009}.

\bibitem[{\citenamefont{BES}(2005)}]{BES:2005ds}
\bibinfo{author}{\bibnamefont{BES}} (\bibinfo{collaboration}{BES})
  (\bibinfo{year}{2005}), \eprint{hep-ex/0506045}.

\bibitem[{\citenamefont{Anselmino and Murgia}(1993)}]{Anselmino:1992rw}
\bibinfo{author}{\bibfnamefont{M.}~\bibnamefont{Anselmino}} \bibnamefont{and}
  \bibinfo{author}{\bibfnamefont{F.}~\bibnamefont{Murgia}},
  \bibinfo{journal}{Phys. Rev.} \textbf{\bibinfo{volume}{D47}},
  \bibinfo{pages}{3977} (\bibinfo{year}{1993}).

\bibitem[{\citenamefont{Lepage and Brodsky}(1980)}]{Lepage:1980fj}
\bibinfo{author}{\bibfnamefont{G.~P.} \bibnamefont{Lepage}} \bibnamefont{and}
  \bibinfo{author}{\bibfnamefont{S.~J.} \bibnamefont{Brodsky}},
  \bibinfo{journal}{Phys. Rev.} \textbf{\bibinfo{volume}{D22}},
  \bibinfo{pages}{2157} (\bibinfo{year}{1980}).

\bibitem[{\citenamefont{Chernyak and Zhitnitsky}(1984)}]{Chernyak:1983ej}
\bibinfo{author}{\bibfnamefont{V.~L.} \bibnamefont{Chernyak}} \bibnamefont{and}
  \bibinfo{author}{\bibfnamefont{A.~R.} \bibnamefont{Zhitnitsky}},
  \bibinfo{journal}{Phys. Rept.} \textbf{\bibinfo{volume}{112}},
  \bibinfo{pages}{173} (\bibinfo{year}{1984}).

\bibitem[{\citenamefont{Ball and Braun}(1996)}]{Ball:1996tb}
\bibinfo{author}{\bibfnamefont{P.}~\bibnamefont{Ball}} \bibnamefont{and}
  \bibinfo{author}{\bibfnamefont{V.~M.} \bibnamefont{Braun}},
  \bibinfo{journal}{Phys. Rev.} \textbf{\bibinfo{volume}{D54}},
  \bibinfo{pages}{2182} (\bibinfo{year}{1996}), \eprint{hep-ph/9602323}.

\end{thebibliography}

\end{document}